\documentstyle[prl,aps,epsf,preprint]{revtex}

\begin{document}
\draft
\title{Thickness dependent magnetotransport in ultra-thin manganite films}
\author{J. Z. Sun and D. W. Abraham}
\address{IBM T. J. Watson Research Center,\\
P. O. Box 218,\\
Yorktown Heights, NY 10598}
\author{R. A. Rao and C. B. Eom}
\address{Department of Mechanical Engineering and Materials Science\\
Duke University \\
Durham, NC 27708}
\date{\today }
\maketitle

\begin{abstract}
To understand the near-interface magnetism in manganites, uniform,
ultra-thin films of La$_{0.67}$Sr$_{0.33}$MnO$_3$ were grown epitaxially on
single crystal (001) LaAlO$_3$ and (110) NdGaO$_3$ substrates. The
temperature and magnetic field dependent film resistance is used to probe
the film's structural and magnetic properties. A surface and/or interface
related dead-layer is inferred from the thickness dependent resistance and
magnetoresistance. The total thickness of the dead layer is estimated
to be $\sim 30~\AA $ for films on NdGaO$_3$ and $\sim 50~\AA $ for films on
LaAlO$_3$.
\end{abstract}
\label{firstpage}
\newpage
One issue in manganite trilayer junction is the premature disappearance of
magnetoresistance (MR) upon temperature increase. The MR in La$%
_{0.67}$Sr$_{0.33}$MnO$_3$ based trilayer junctions disappears above $150\,K$, 
well below the electrode's Curie temperature of $360~K$
\cite{142,224,293,191,390,98165,98191}. Recently, spin-resolved photoemission
measured the spin-polarization of surface electronic density-of-state of 
epitaxial thin films of La$_{0.67}$Sr$_{0.33}$MnO$_3$ \cite{98190,98078}. 
It revealed, as a function of increasing temperature, a more 
rapid decrease of the surface spin polarization
than the film's overall magnetization. Both transport and photoemission 
experiments suggest the
possible existence of a surface dead-layer with depressed magnetic order at
elevated temperatures. Questions remain as to the depth of this dead-layer,
and its physical origin. In this paper, we experimentally establish some
estimates about the dead-layer thickness. The approach is to study the thickness
dependent electrical transport, including resistance and magnetoresistance,
in ultra-thin epitaxial films of La$_{0.67}$Sr$_{0.33}$MnO$_3$ (LSMO).
To estimate the effect of lattice strain, films on two types of substrates
are compared --- those grown on (001) LaAlO$_3$(LAO)  and on (110) NdGaO$_3$
(NGO). The film thickness range covered by this study is between $15\,\AA $
and $240\,\AA $.

Films were grown epitaxially using laser ablation from a stoichiometric La$%
_{0.67}$Sr$_{0.33}$MnO$_3$ target. A Nd-YAG laser was used for ablation, 
operating in frequency-tripled mode at $355~nm$, with a pulse energy of  
$140\,mJ/pulse$ at $10\,Hz$ repetition. The deposition rate was
about $0.6\,\AA /\sec $. A series of films were made in the thickness range
between $15~\AA $ and $240~\AA $. Two substrates, one LAO and
one NGO, each $1\,cm\times 2\,mm\times 0.5\,mm$ in size, were loaded
side-by-side for simultaneous deposition at each thickness. The substrate
holder's temperature during growth was $750\,C$, and the growth ambient
consisted of $300\,m\tau $ of oxygen at a flow rate of $50\,sccm$.

The films were subsequently coated with four in-line contact pads, $1~mm$
diameter and $2.2~mm$ apart from each other, for transport measurement.
These contact pads, usually silver or gold, about $1000~\AA $ thick, were
sputter deposited through a stencil mask. The four-point resistance thus 
measured was assumed to be the resistance-square $R_{\Box }$. The film
thickness was deduced from deposition time normalized to calibration
runs.

A summary of the temperature dependent film resistivity, $\rho (T)$, is shown
in Fig.\ref{figfile1}(a). An increase of resistivity is seen as
the film thickness is decreased, for films both
on NGO and on LAO. Films on LAO show faster increase
in resistivity upon thickness reduction. At $15~\AA $, the $\rho (T)$ for
film on NGO\ just begins to show a negative temperature slope at $14~K$, with 
its $\rho (T)$ peak temperature moved down to around $240~K$\ from the bulk value of 
$360K$. The $\rho (T)$ for the $15~\AA$ film on LAO, on the other hand, already 
shows a pronounced 
resistivity rise for $T<150~K$. The increase of resistivity with decreasing film
thickness indicates the presence of a dead-layer with reduced conductivity. 
This can be seen more clearly in the thickness
dependence of the film's {\it total} conductance $G$ (defined as $G=1/R_{\Box}$,
where $R_{\Box} $ is the film's 4-probe-measured resistance).

Fig.\ref{figfile1}(b) shows the thickness dependence of $G$ at $T=14~K$,
the lowest temperature the film resistance was measured down to. A linear
thickness dependence of $G$ is expected. If the films were uniform, the
line should intercept the horizontal axis at zero thickness. Data in Fig.\ref
{figfile1}(b) shows a linear thickness dependence for films thicker
than $60~\AA$, but with a
finite intercept. This intercept represents the total thickness of
the dead-layer(s), if these dead-layers are much less
conducting than the center part of the film. The total dead-layer thickness
according to this estimate is around $30\,\AA $ for films grown on NGO, and
around $50\,\AA $ for films on LAO. This thickness estimate combines the
contribution of the surface dead-layer and the film-substrate interface
dead-layer. It is not yet clear whether the thicknesses of these two types
of dead-layers are the same for a given film.

Next we rule out two other possibilities that could contribute to a 
thickness-dependent film resistivity, namely the film's strain variation,
and the incomplete film coverage over the substrate during initial growth.

Fig.\ref{figfile2} shows results of X-ray diffraction measurements 
on lattice parameters of the films. Both out-of-plane and in-plane 
lattice parameters were measured, using normal and grazing incident 
X-ray diffraction\cite{Eom}. The film's unit cell dimensions deviate 
from their bulk values for the entire thickness range, presumably due
to the combined effect of epitaxial strain and oxygen defects. The deviation 
is greater for films on LAO\ than for films on NGO. A thickness-dependent 
change of lattice parameter is seen. This change, $<0.76\%$ for 
out-of-plane, $<1.2\%$ for in-plane lattice in the case of film on LAO, is small
compared to the difference in corresponding lattice parameters between films 
grown on LAO and NGO.

The main contribution to resistivity's thickness dependence is not from film
strain variation. This can be seen by comparing the strain-field difference
and resistivity difference for films grown on different substrates. For the $%
240\,\AA $ film on LAO\ and NGO, their out-of-plane lattice constants
differ from bulk value by $2.12\%$ and $0.83\%$, respectively. This is a
difference in strain-field of over a factor of $2.5$ in the out-of-plane direction.
Yet their averaged resistivity at $14~K$ differ only by a factor of $1.3$, with
dead-layer effects included. Thus, a relatively small
strain variation of $0.2\%\sim 0.5\%$ over the thickness range is not
expected to cause much variation to the resistivity. The observed 
thickness-dependent resistivity varies over an order of magnitude, 
and can not be explained away by strain variation.

The continuous coverage of such thin films over the substrate surface was
confirmed using atomic force microscopy (AFM). Fig.\ref{figfile3} shows a
morphology image of two $30\,\AA $ films, one on LAO, another
on NGO. From these images the film's surface roughness is estimated to be
about $5~\AA $ peak-to-peak in both cases, indicating continuous film
coverage without large voids. 

With strain effect and initial coverage issues
resolved, we thus conclude that the thickness-dependent resistivity 
is most likely caused by the presence of a surface and/or interface dead-layer.

The MR of the films also show thickness and
substrate dependence. Fig.\ref{figfile4} shows MR {\it vs.} temperature
for films on NGO and LAO at different thicknesses. Two
types of behaviors are distinguishable. In the first type the MR
rises monotonically as a function of temperature. This is
what normally expected, since the Curie temperature of bulk LSMO is around $%
360K$ which is above the maximum measurement temperature of $300~K$. This
normal behavior is associated with thicker films. The second type of MR {\it vs.}
temperature shows a peak between $100 \sim 200~K$. This only 
appears in very thin films ($15\,\AA $ film on NGO, $30$ and $60\,\AA $
film on LAO). It is the signature MR behavior of the dead-layer. The peak 
temperature in MR could be used as an empirical estimate\cite{Schiffer1} to
the Curie temperature, $T_c$ of the dead-layer, which thus comes to be around  
$100 \sim 200~K$.

In conclusion, we have obtained an experimental estimate of the total
thickness of the dead-layers associated with the surface and film-substrate
interface in epitaxial films of LSMO on NGO and LAO substrates. The total
dead-layer thickness in these two cases are estimated to be about $30~\AA $
and $50~\AA $, respectively. The dead-layer has a magnetoresistance peak
temperature of around $100$ to $200~K$, and is likely to have a magnetic
Curie temperature in the same range.

JZS wishs to thank Pat Mooney for some early X-ray analysis.
CBE\ would like to acknowledge support from NSF Grant No. DMR9802444 and
from NSF's Young Investigator Award.


\begin{references}

\bibitem{142}
J.~Z. Sun, W.~J. Gallagher, P.~R. Duncombe, L. Krusin-Elbaum, R.~A. Altman, A.
  Gupta, Y. Lu, G.~Q. Gong, and G. Xiao, Appl. Phys. Lett. {\bf 69},  3266
  (1996).

\bibitem{224}
J.~Z. Sun, L. Krusin-Elbaum, A. Gupta, G. Xiao, P.~R. Duncombe, and S.~S.~P.
  Parkin, IBM J. of Res. and Dev. {\bf 42},  89  (1998).

\bibitem{293}
J.~Z. Sun, L. Krusin-Elbaum, P.~R. Duncombe, A. Gupta, and R.~B. Laibowitz,
  Appl. Phys. Lett. {\bf 70},  1769  (1997).

\bibitem{191}
Y. Lu, X.~W. Li, G.~Q. Gong, G. Xiao, A. Gupta, P. Lecoeur, J.~Z. Sun, Y.~Y.
  Wang, and V.~P. Dravid, Phys. Rev. B {\bf 54},  R8357  (1996).

\bibitem{390}
M. Viret, M. Drouet, J. Nassar, J.~P. Contour, C. Fermon, and A. Fert,
  Europhysics Letters {\bf 39},  545  (1997).

\bibitem{98165}
J.~Z. Sun and A. Gupta, Annu. Rev. Mater. Sci. {\bf 28},  45  (1998).

\bibitem{98191}
J.~Z. Sun, Phil. Trans. R. Soc. Lond. A {\bf 356},  1693  (1998).

\bibitem{98190}
J.-H. Park, E. Vescovo, H.-J. Kim, C. Kwon, R. Ramesh, and T. Venkatesan, Phys.
  Rev. Lett. {\bf 81},  1953  (1998).

\bibitem{98078}
J.-H. Park, E. Vescovo, H.-J. Kim, C. Kwon, R. Ramesh, and T. Venkatesan,
  Nature {\bf 392},  794  (1998).

\bibitem{Eom}
R. A. Rao, D. Lavric, T. K. Nath, C. B. Eom, L. Wu and F. Tsui, Submitted 
to Appl. Phys. Lett. (1998).

\bibitem{Schiffer1}
See, for example: P. Schiffer, A. P. Ramirez, W. Bao, and S-W. Cheong, Phys. Rev. Lett.
{\bf 75}, 3336 (1995).

\end{references}

\begin{figure}
\caption{(a)Temperature-dependent resistivity for films of different
thicknesses. Resistivity increases for thinner films. 
(b)Thickness dependence of the total conductance of films at 14K. 
For films that are metallic, $R(T)$ is flat enough in this temperature
region to consider this value representative of the residual resistance. 
The thickness dependence of conductance is linear, but with a
finite intercept on the horizontal axis, suggesting the existence of
an electrically less conducting dead-layer. From the intercept the total
thickness of the dead layers can be estimated. For films on NGO, it is about 
$30~\AA $; on LAO, about $50~\AA $.}
\label{figfile1}
\end{figure}

\begin{figure}[tbp]
\caption{Lattice parameters' thickness dependence as measured using a
4-circle X-ray diffractometer.  Error bars reflect the diffraction peak-width.
A small variation of film lattice constant
is seen to occur when the film reaches a thickness of around $60$ to 
$120~\AA $. The small 
increase of out-of-plane lattice constant {\it vs.} thickness for films on
NGO is not well understood. The presence of large substrate 
peaks nearby may account for some variation in 
diffraction peak position (and therefore lattice constant) for the 
very thin films of $15~\AA \sim 30~\AA$.}
\label{figfile2}
\end{figure}

\begin{figure}[tbp]
\caption{AFM images of two $30~\AA $ films, one grown on NGO (a), the other
on LAO (b). Both films have similar peak-to-peak surface roughness,
estimated to be less than $5~\AA $, excluding large particles believed to be
laser-ablation-related particulates. This suggests continuous film coverage.}
\label{figfile3}
\end{figure}

\begin{figure}[tbp]
\caption{Thickness dependent evolution of magnetoresistance for films
on NGO(a) and on LAO(b). Magnetoresistance are evaluated with a field
sweeping of $\pm 3~kOe$. A low-temperature peak in
magnetoresistance is seen for the very thin films. The threshold for this
behavior is $<60~\AA $ for films on LAO and $\sim 15~\AA $ for films
on NGO, consistent with estimates of the dead layer thicknesses of data from
Fig.\ref{figfile1}(b).}
\label{figfile4}
\end{figure}

\newpage
\epsfxsize=6in \epsfbox{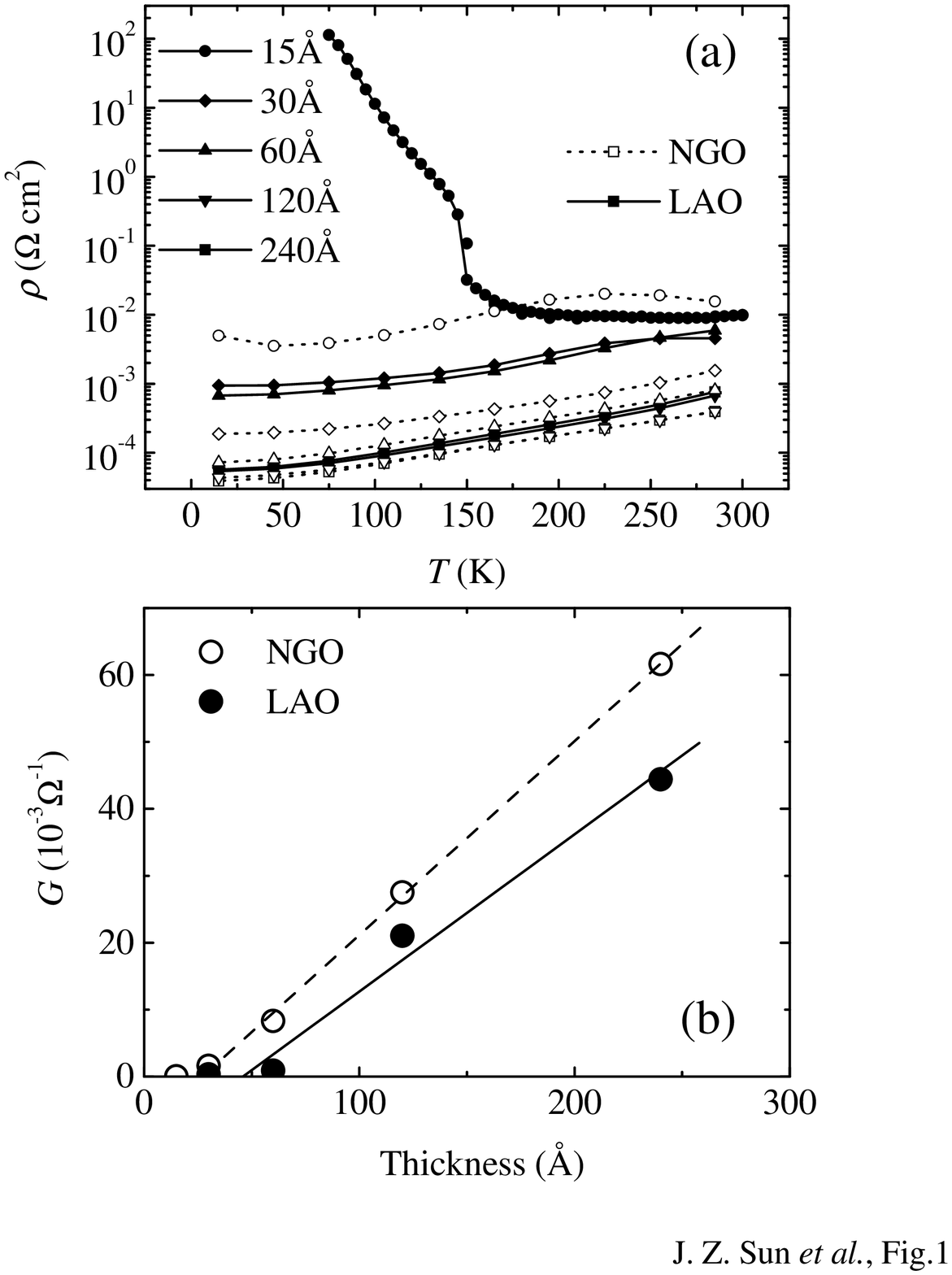} \newpage
\epsfxsize=6in \epsfbox{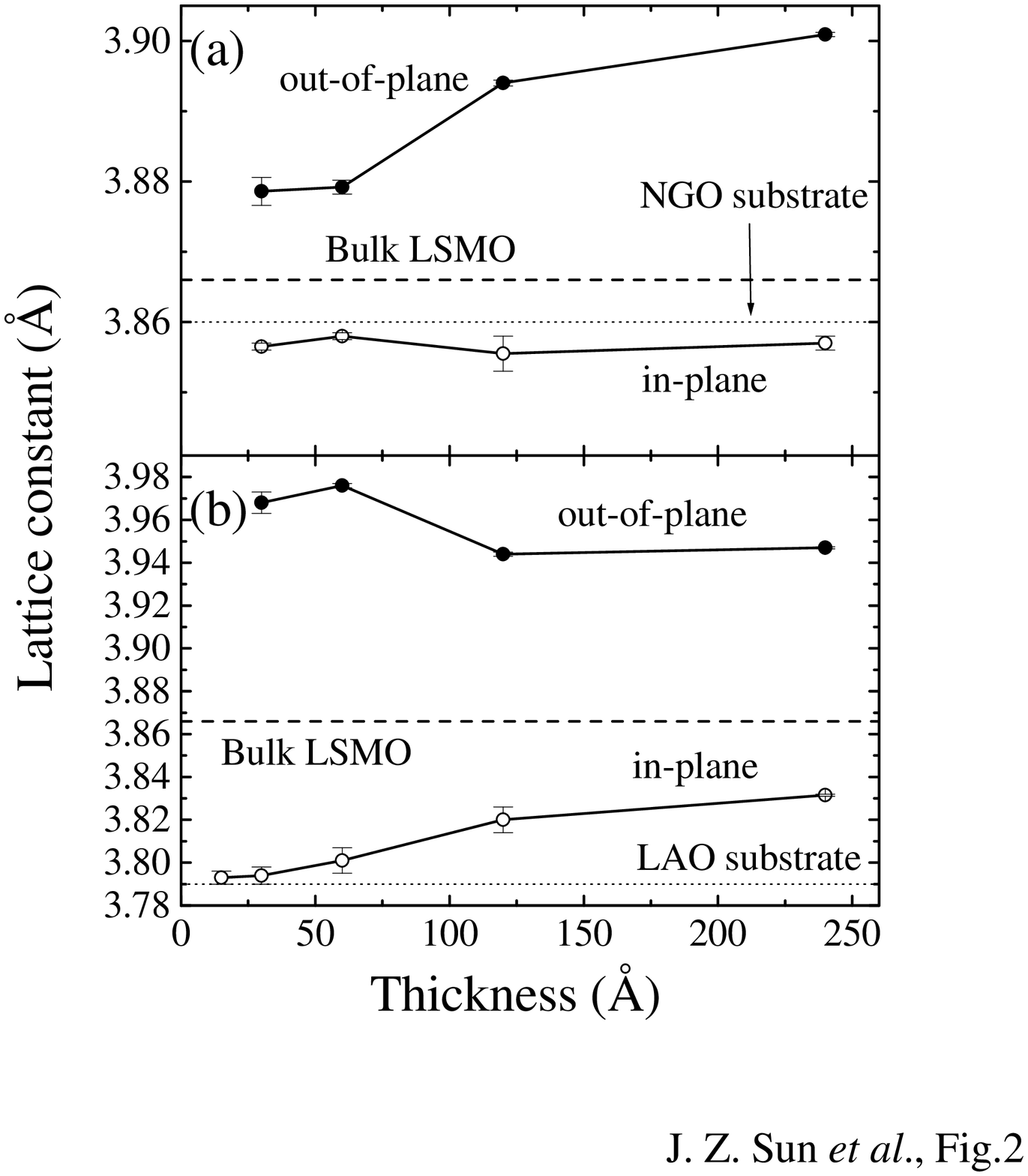} \newpage
\epsfxsize=6in \epsfbox{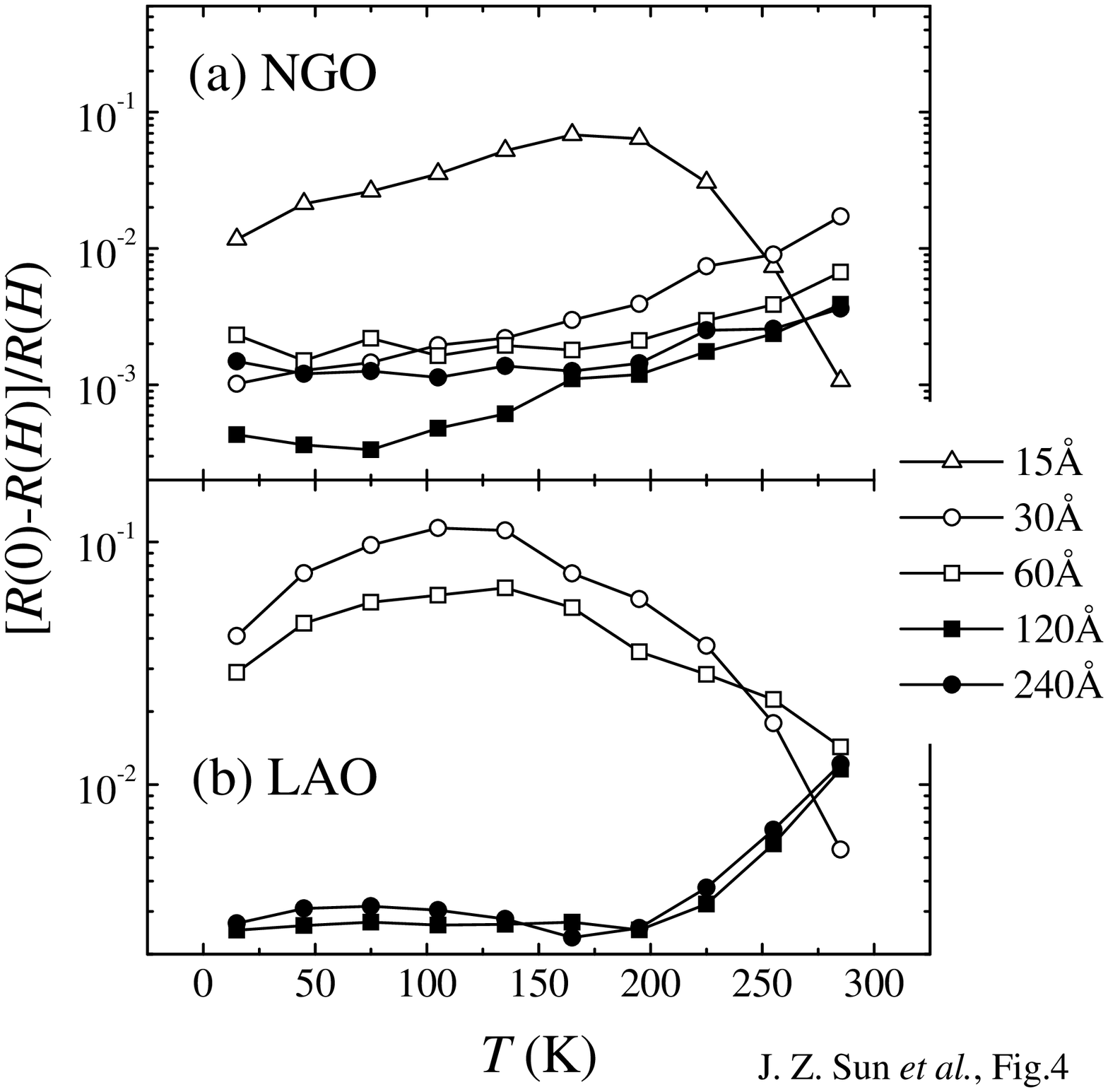}
\end{document}